\documentclass[10pt]{iopart}
\usepackage{psfig}
\newcommand{\be}{\begin{equation}}
\newcommand{\ee}{\end{equation}}

\newcommand{\bea}{\begin{eqnarray}}
\newcommand{\eea}{\end{eqnarray}}

\begin{document}

\title[Enhanced Strange Particle Yields - Signal of a Phase of Massless Particles ?]{
Enhanced Strange Particle Yields -\\ 
Signal of a Phase of Massless Particles ?
}

\author{Sven Soff$\,^*\,\sharp$\ddag, D.~Zschiesche\ddag, 
M.~Bleicher\ddag, C.~Hartnack$\parallel$,
M.~Belkacem$^+$,
L.~Bravina\P, E.~Zabrodin\P, S.A.~Bass\S,\\
H. St\"ocker\ddag, and W. Greiner\ddag}

\address{$^*$Brookhaven National Laboratory, Physics Department, 
P.O. Box 5000, Upton, New York 11973-5000, USA}  
\address{$\sharp$ Gesellschaft f\"ur Schwerionenforschung, Postfach 110552, D-64220 Darmstadt, Germany}
\address{\ddag Institut f\"ur Theoretische Physik der
J.W.Goethe-Universit\"at\\
Postfach 111932, D-60054 Frankfurt am Main, Germany}
\address{$\parallel$ SUBATECH, \'Ecole des Mines, F-44072 Nantes, France}
\address{$^+$School of Physics and Astronomy, University of Minnesota, 
Minneapolis, MN 55455, USA}
\address{\P Institut f\"ur Theoretische Physik, Universit\"at T\"ubingen\\
        $^{~}$Auf der Morgenstelle 14, D-72076 T\"ubingen}
\address{\S Department of Physics, Duke University, Durham, NC 27708, USA\\}

\begin{abstract}
The yields of strange particles are calculated with the 
UrQMD model for p,Pb($158\,A$GeV)Pb collisions 
and compared to experimental data. The yields are enhanced 
in central collisions if compared to proton induced or 
peripheral Pb+Pb collisions. The enhancement is due to 
secondary interactions. Nevertheless, only a reduction of 
the quark masses or equivalently an 
increase of the string tension provides an adequate description of 
the large observed enhancement factors (WA97 and NA49).
Furthermore, the yields of 
unstable strange resonances as the $\Lambda^*$ or 
the $\phi$ are considerably affected by hadronic rescattering 
of the decay products.  
\end{abstract}

\section{Introduction}
Strange particle yields are most interesting and useful probes 
to examine excited nucelar matter 
\cite{raf8286,soff99plb,senger99,JPG,stock99,and98a,sikler99,cgreiner00} 
and to detect the transition of (confined)
hadronic matter to quark-gluon-matter (QGP).
The relative enhancement of strange and especially multistrange 
particles and the hadron ratios in central 
heavy ion collisions with respect to peripheral or proton 
induced interactions have been suggested as a signature 
for the 
transient existence of a
QGP-phase \cite{raf8286}. 
The production mechanism of $s\overline{s}$ pairs via gluon fusion
($gg \rightarrow s \overline{s}$) \cite{raf8286} 
in an equilibrated (gluon rich) plasma phase
should allow for equilibration times much shorter (a few fm/c) than 
a thermally equilibrated hadronic fireball of $T\sim 160\,$MeV.

Measurements by the WA97 and the NA49 collaborations
clearly demonstrated the relative enhancement
of the (anti-)hyperon yields ($\Lambda$, $\Xi$, $\Omega$)
in Pb-Pb collisions as
compared to p-Pb collisions \cite{and98a,sikler99}.
The observed enhancement increases with the strangeness content ($|S|=1,2,3$)
of the probe under
investigation \cite{and98a,sikler99}.
For the ($\Omega^- + \overline{\Omega^-}$)-yield the enhancement factor
is as large as 15 !

Here, 
strangeness production is investigated within 
a microscopic transport model:
hadronic and string degrees of freedom are employed in
the Ultrarelativistic Quantum Molecular Dynamcis model
(UrQMD) \cite{bas98}.
The strange baryon yields for Pb$(158\,A\,{\rm GeV})$Pb
collisions are computed
vs. centrality and for p$(158\,{\rm GeV})$Pb collisions.
The observed total yields of $\Lambda$'s, $\Xi$'s and $\Omega$'s are
well described in the p-Pb case by the present model.
Strangeness enhancement is predicted in the present
calculation for Pb-Pb due to rescattering.
However, for central Pb-Pb collisions the
experimentally observed hyperon yields are underestimated by
the present calculations.
The discrepancy to the data increases strongly with the
strangeness content of the hadron.

An ad hoc overall increase of the color electric field
strength (effective string tension of
$\kappa=3\,{\rm GeV/fm}$), or, equivalently, a reduction of the constituent
quark masses to the current quark masses enhances the hyperon yields to
the experimentally observed high values.

Enhancement factors of $\approx 1.5 (2)$ for $\Lambda$'s,
$\approx 2 (6)$ for $\Xi^-$'s,
and $\approx 5 (13)$  for $\Omega^-$'s
are obtained at midrapidity. The values in brackets are the results
of the reduced masses/enhanced
string tension calculations.
The enhancement depends strongly on rapidity. 

The reconstructed yields of strange resonances are shown 
to be strongly affected by the rescattering of their decay 
products. 
$\phi$ yields are reduced by $\approx 25\%$ due to 
kaon rescattering, 
$\Lambda^*(1520)$ yields are even reduced by 
$\approx 50\%$! 
As a consequence the observed differences of the 
inverse slope parameters $T$ of $\phi$ mesons (NA40 vs.\ NA50) 
can be explained qualitatively. 

Complementary, also the late resonance production via 
hadronic rescattering processes plays an important role, e.g.\ 
K$^+$K$^- \rightarrow \phi$.

\section{The UrQMD model - one test}
The Ultrarelativisitc Quantum Molecular Dynamics model (UrQMD) 
has been developed to simulate heavy ion
collisions in a wide energy range (from GSI/SIS to BNL/RHIC).
A detailed documentation of the underlying concepts of the model and
comparisons to experimental data are available in 
\cite{bas98,soff00}.
The two main constituents of the model are on the one hand 
the particles that are incorporated and propagated and on the 
other hand the cross sections for the various collision processes. 
On the order of one hundred known ground and excited baryon and meson states 
are in the model in addition to the higher mass states modeled 
via string degrees of freedom. 
The string tension $\kappa$ is set to $\kappa=1\,$GeV/fm \cite{LUND}.
Discussing the strangeness production via string excitation 
and fragmentation \cite{LUND} 
the production probability of a $s\bar{s}$ pair is reduced as compared to
$u\bar{u} / d\bar{d}$-pairs according to the
Schwinger formula \cite{schwing51}
\begin{equation}
\gamma_s=\frac{P(s\bar{s})}{P(q\bar{q})}=\exp
\left(- \frac{\pi (m_s^2-m_q^2)}{2\kappa}\right)\,.
\label{eq1}
\end{equation}
Here, the masses of the strange $m_s$ and the up/down $m_q$ quarks enter.
The constituent quark mass values $m_q=0.3\,$GeV and $m_s=0.5\,$GeV 
yield a strangeness suppression factor $\gamma_s \approx 0.3$. 

One of several tests of the model is to demonstrate 
that the particle production 
yields are reproduced reasonably 
in elementary collisions. 
Fig.1 and Fig.2 show the comparisons of $\pi^{\pm},\,$K$^{\pm},\,
\Lambda$ and $\overline{\Lambda}$ yields to 
experimental data in proton proton collisions as a function of 
the center of mass energy. 
The shown agreement over a large energy range 
supports the chosen set of model parameters, e.g.\ the parameters for the 
string decay, and provides a good starting point 
to investigate nucleus nucleus collisions.   
The $\overline{\Lambda}$'s might be slightly overpredicted by the 
model calculations but in this case the data show uncertainties 
as well (Fig.2).

\parbox{6.5cm}{
\hspace*{-1cm}\psfig{figure=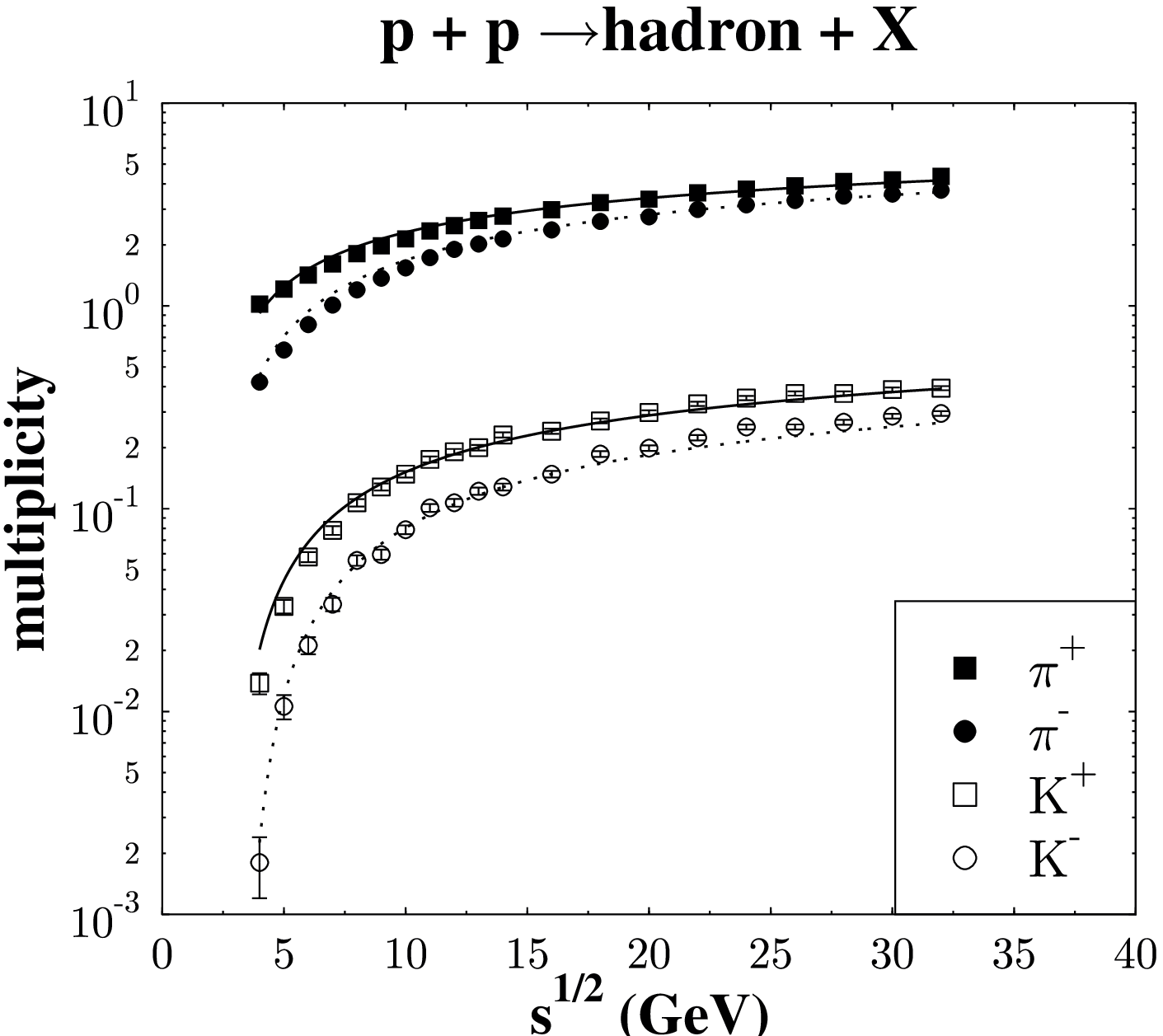,width=7cm}
{\hspace*{-.5cm}\small \mbox{     }Figure 1: Multiplicities of $\pi^{\pm}$ and 
K$^{\pm}$ in pp\\ 
\hspace*{-.5cm}collisions as a function of the center of mass\\ 
\hspace*{-.5cm}energy 
$\sqrt{s}$. The symbols show results of the\\
\hspace*{-.5cm}
UrQMD calculations 
and lines represent\\ \hspace*{-.5cm}parametrizations 
of experimental data.}
}
\parbox{6.5cm}{
\hspace*{-1.5cm}
\psfig{figure=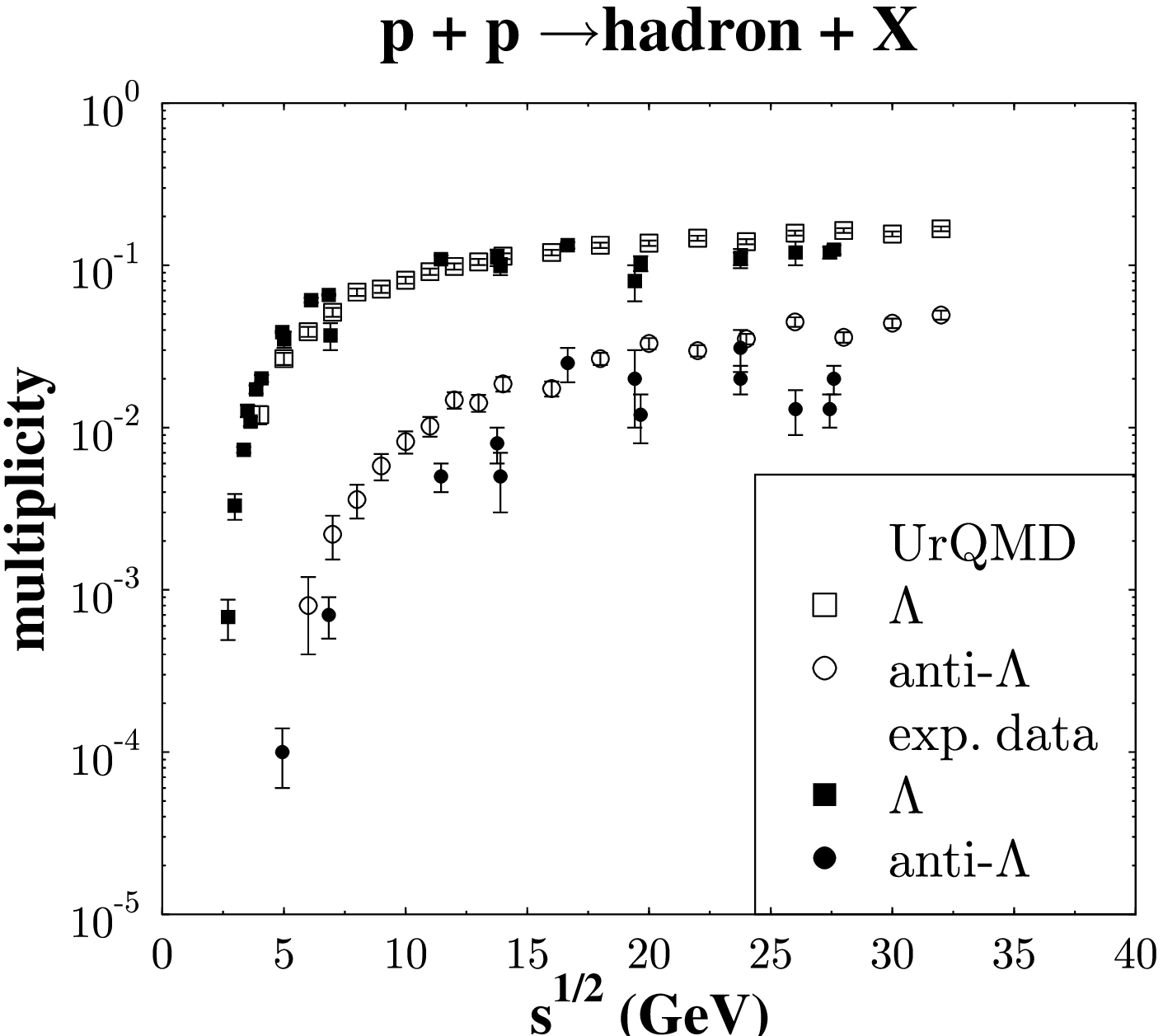,width=7.2cm}    
{\small \mbox{    }Figure 2: Multiplicities of $\Lambda$ and    
$\overline{\Lambda}$ in pp collisions as a function 
of the center of mass energy    
$\sqrt{s}$. Open symbols show UrQMD results, 
full symbols are experimental data.}
}

\section{Comparison to data and reduced quark masses}

Fig.3 shows (anti-)hyperon yields in Pb($158\,A$GeV)Pb 
collisions (as a function of the number of participants) and in 
pPb collisions \cite{soff99plb}. 
While the agreement between data of WA97 (stars) and 
UrQMD (open symbols) seems to be reasonable in pPb collisions 
discrepancies are observed in central collisions that increase 
with the strangeness content of the probe. 
Motivated by chiral symmetry restoration the quark masses may 
be reduced that enter in (1) thus enhancing the strangeness suppression 
factor $\gamma_s$ \cite{soff99plb}. 
A reduction of the constituent quark masses 
to the current quark masses $m_q=10\,$MeV and $m_s=230\,$MeV 
yields $\gamma_s \approx 0.65$ \cite{soff99plb}. 
Similarly, the superposition of several strings 
may increase the color electric field strength \cite{biro84,sor92}, 
thus leading 
to an effectively enhanced string tension. 
Enhancing the string tension to $\kappa = 3\,$GeV/fm 
yields $\gamma_s\approx 0.65$, too \cite{soff99plb}.
The reduced masses/enhanced string tension UrQMD calculations
enhance the calculated yields to the data as shown in Fig.3 
(full symbols).  
Since the asymmetric p+Pb collisions are not ideally suited 
for the comparison to central Pb+Pb collisions 
(e.g.\ asymmetric rapidity distributions \cite{soff99plb})  
the enhancement factors $E_Y$ of the hyperons $Y$ are determined by 
$E_Y=(Y/\pi^-)_{\rm cent}/(Y/\pi^-)_{\rm peri}$ in Pb+Pb. 
As shown in Fig.4 they reach their maximum at midrapidity. 
Open symbols are the enhancement factors for the UrQMD 
calculations while full symbols show the results if 
the reduced masses/enahnced string tension calculations are 
taken into account for the central collisions. 
For the $\Omega$'s the enhancement factor can be as large 
as $15$ (only for the reduced masses/enhanced string tension case). 
A microscopic analysis shows that most of the $\Omega$'s 
are produced in secondary meson baryon collisions ($\approx 80\%$) 
(e.g.\ meson $\Lambda$, meson $\Xi$)   
whereas baryon baryon collisions contribute on the order of 
$20\%$ (e.g.\ nucleon nucleon, nucleon $\Lambda$, 
nucleon $\Xi$). 
About $15\%$ of the produced $\Omega$'s are again absorbed 
(e.g.\ by $\Omega K,\,\Omega\overline{B}$). 
  
\parbox{7.8cm}{
\psfig{figure=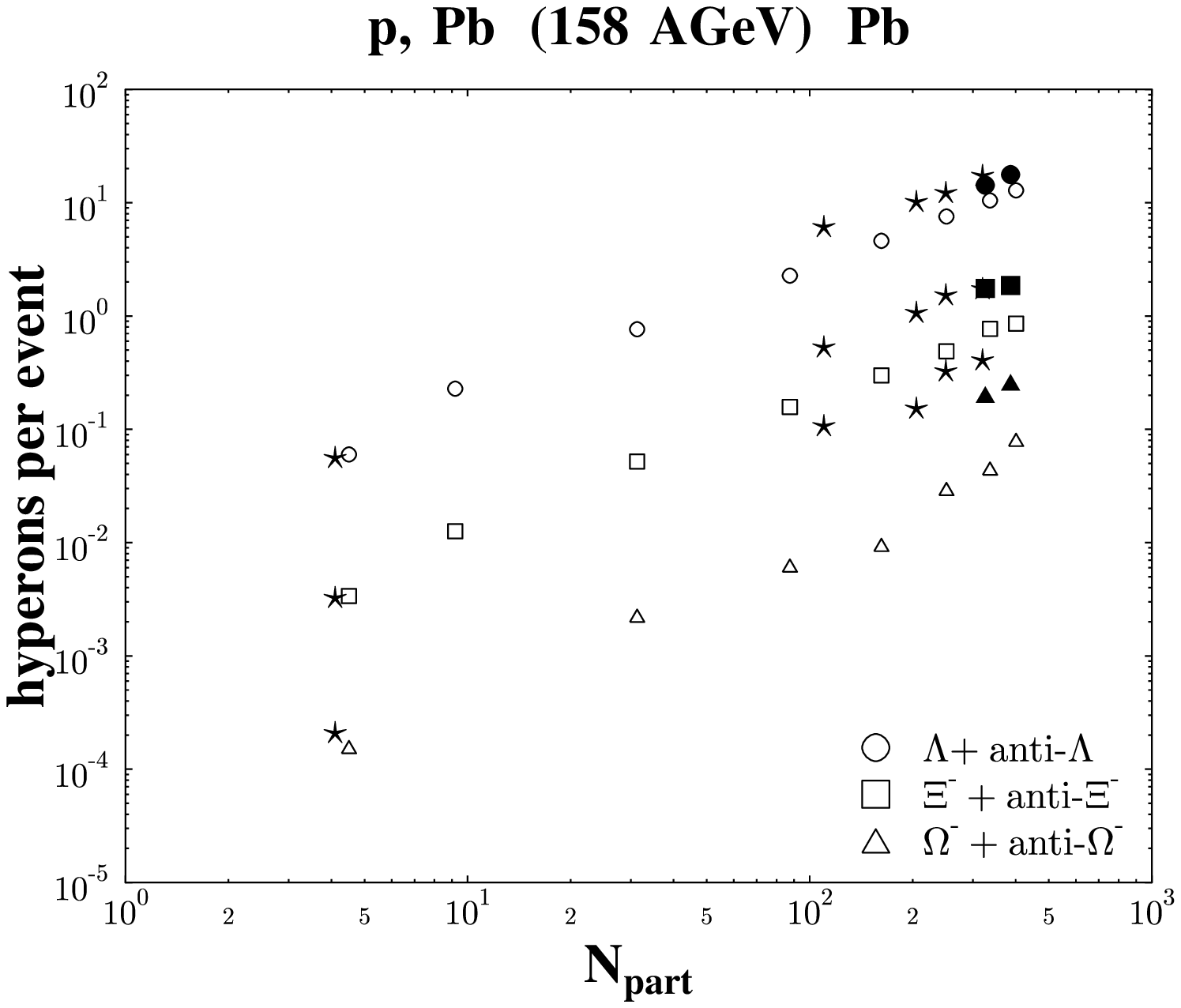,width=7.cm}    
}
\parbox{5cm}{{\small Figure 3: Multiplicities of \mbox{(anti-)hyperons} 
per event at midrapidity 
$|y-y_{\rm cm}|<0.5$ as a function of the number of 
participants $N_{\rm part}$ 
for Pb-Pb and p-Pb collisions at $158\,A$GeV. 
Stars are data of WA97, open symbols are UrQMD results and full symbols are 
UrQMD results with reduced masses or enhanced string tension.
}
}

\parbox{7.3cm}{
\psfig{figure=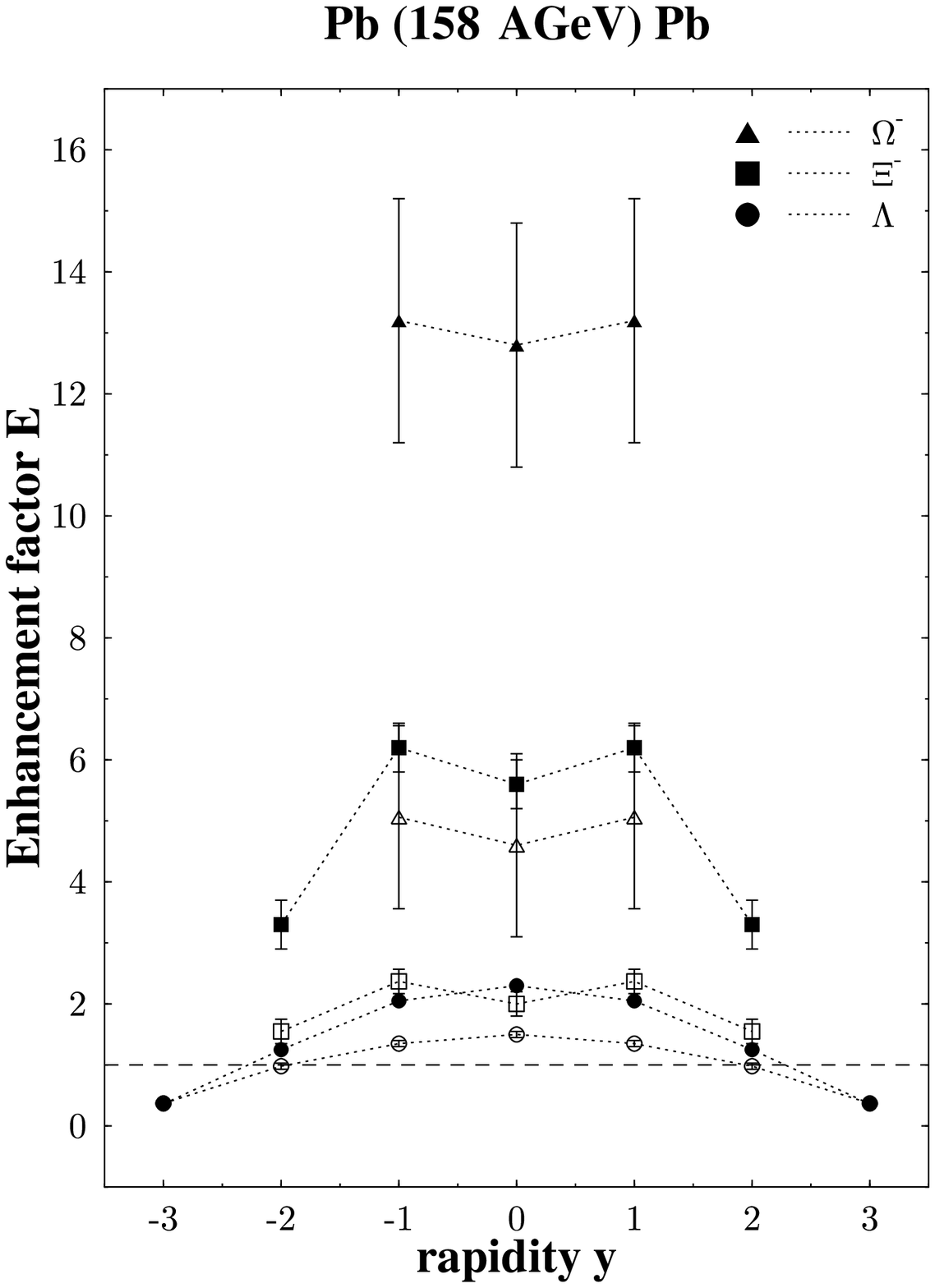,width=7.cm}    
}
\parbox{5cm}{{\small Figure 4: Enhancement factors $E_Y$ for 
$\Lambda$'s (circles), $\Xi$'s (squares) and $\Omega$'s (triangles) 
as a function 
of rapidity in Pb($158\,A$GeV)Pb collisions.
Open symbols show UrQMD calculations, full symbols represent 
results when the 
reduced masses/enhanced string tension calculations 
are taken for the central collisions. 
$E_{\Omega^-}$ may reach a value of 15! 
Strangeness suppression is predicted for $\Lambda$'s at 
target/projectile rapidity.
}
}

In order to support the line of arguments that lead to the reduction 
of quark masses Fig.5 shows the baryonic particle densities 
as a function of temperature 
as obtained from a chiral $SU(3) \times SU(3) $ $\sigma-\omega$ model 
\cite{Zschiesche:2000gf}.
The model selfconsistently predicts a phase transition for 
$T \approx 150\,$MeV and $\mu =0$ \cite {Zschiesche:2000gf}. 
At this point the masses of the baryons 
drop drastically and this leads to the strong enhancement 
of baryonic particle 
(and antiparticle) densities.

\parbox{7.3cm}{
\hspace*{-1cm}
\psfig{figure=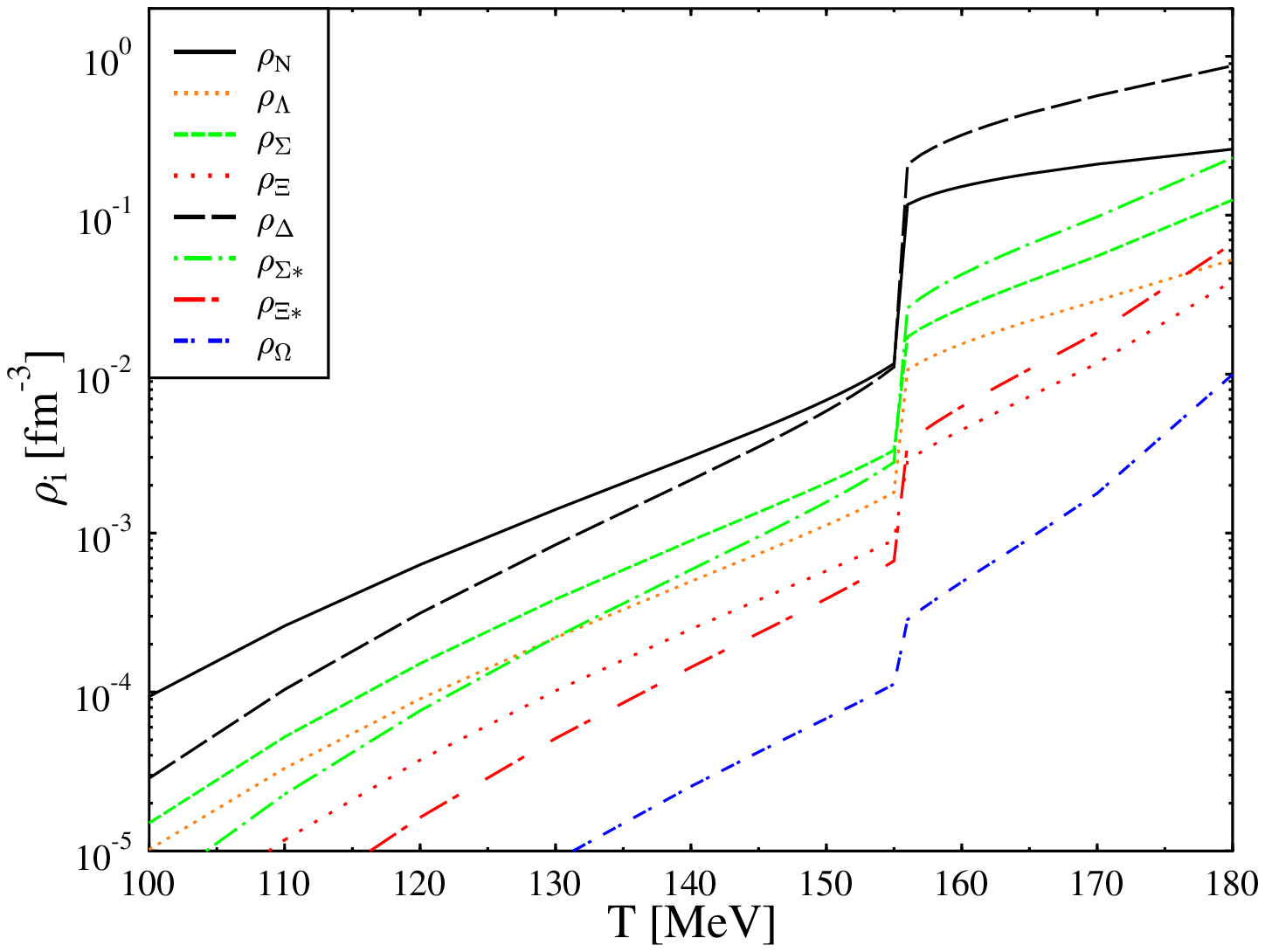,width=8cm}  
}
\parbox{5cm}{{\small Figure 5: 
Baryonic particle densities as a function of temperature for vanishing 
chemical potential in a chiral SU(3)$\times$SU(3) $\sigma$-$\omega$ model. 
Due to the phase transition at $T\approx 150\,$MeV 
and the strong decrease of the 
baryonic masses, the particle and antiparticle densities strongly increase 
around the phase transition region. 
}
}

\section{Strange resonance yields - $\Lambda^*(1520)$ and $\phi$}
In this section we briefly discuss the production 
of the $\Lambda^*(1520)$ resonance and the $\phi$ meson 
and how the reconstruction of these particles by their 
hadronic decay products is distorted due to 
rescattering. 

\parbox{7.3cm}{
\psfig{figure=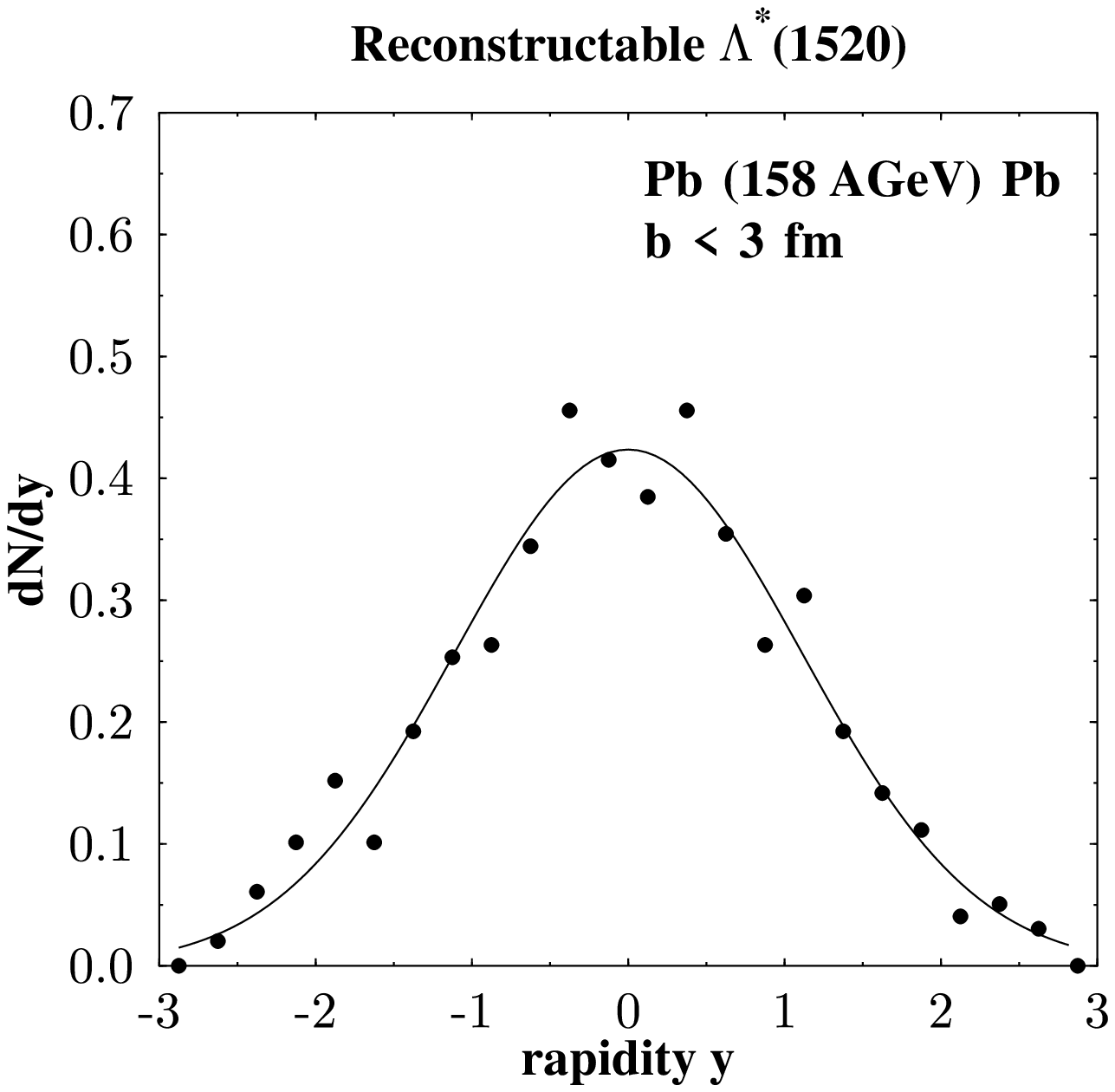,width=7.cm}    
}
\parbox{5cm}{{\small Figure 6: Rapidity distribution $dN/dy$ of 
reconstructable $\Lambda^*(1520)$'s, i.e. $\Lambda^*(1520)$'s whose 
decay products do not rescatter, in central 
$(b<3\,$fm) Pb($158\,A$GeV)Pb collisions as predicted by the UrQMD model. 
The total number of $\Lambda^*(1520)$ decays is about 
twice the number of reconstructable $\Lambda^*(1520)$'s.}
}

The $\Lambda^*(1520)$ ($I(J^P)=0(\frac{3}{2}^-)$) is currently investigated 
by the NA49 collaboration \cite{markert00}
in Pb($158\,A$GeV)Pb collisions. Its free width is $\Gamma\approx15.6\,$MeV 
corresponding to $\tau\approx 12.6\,$fm/c. 
The main decay channels are $N \overline{K} (45\%),\, 
\Sigma\pi (42\%)$, and $\Lambda\pi\pi(10\%)$.  
Especially the $\Lambda^*(1520)\rightarrow p K^-$ decay is suited for 
the reconstruction by invariant mass distributions. 
The UrQMD calculations predict $\approx 2.4$ $\Lambda^*(1520)$ decays 
per central Pb+Pb collision (integrated over all decay channels). 
If this is multiplied with the branching ratio to the $pK^-$ channel 
($22.5\%$) 
and the experimental acceptance ($\approx 50\%$) 
an experimental detection probability of about 
0.25 $\Lambda^*(1520)$ per event is predicted. 
However, subsequent collisions of the 
decay products (e.g.\ $K^-p$) have not been taken into account 
for this estimate. They destroy the signal in the 
invariant mass spectra, thus lowering the observable yield.     
Fig.5 shows the rapidity distribution of those 
$\Lambda^*(1520)$ whose decay products do not suffer 
subsequent collisions. 
Its width is $\sigma_{\rm FWHM} \approx 2.5$. 
Integrating this rapidity distribution yields 
a value of $\approx 1.2\,\Lambda^*(1520)$ per event. 
As a consequence, only half of the produced 
$\Lambda^*(1520)$ could be ideally reconstructed by 
their decay products $x$ and $y$ 
\begin{equation}
\frac{\Lambda^* \rightarrow (xy)_{\rm escape}}
{\Lambda^* \rightarrow (xy)_{\rm all}}\le 0.5\,.
\end{equation}
If an enhancement of the $\Lambda^*$'s due to 
a reduction of the quark masses/enhanced 
string tension is taken into account the above 
yields increase by a factor 1.5, corresponding 
to the enhancement factors of K's and $\Lambda$'s 
with the same strangeness content.

Similarly, $\phi$ meson yields and 
their reconstruction via $\phi \rightarrow K^+K^-$ 
are affected due to rescattering of the 
kaons. In this case the ratio of 
reconstructable to total $\phi$ decays is larger than 
in the $\Lambda^*(1520)$ case (mostly due to 
the longer lifetime of the $\phi$)
\begin{equation}
\frac{\phi \rightarrow (xy)_{\rm escape}}
{\phi \rightarrow (xy)_{\rm all}}\le 0.75\,.
\end{equation}
However, this approximate $25\%$ decrease of the observable yield 
improves the discrepancies between data and 
microscopic model predictions (e.g.\cite{soff99plb}).
The rapidity distributions show that the decay products are 
preferentially rescattered around midrapidity thus 
broadening the reconstructable $\phi$ rapidity 
distribution. Complementary the decay products are rescattered 
preferentially at low transverse momenta thus 
pretending a larger inverse slope parameter, as also shown 
in \cite{johnson00}. 
This corresponds to the different measured inverse slope 
parameters of the NA49 ($\phi\rightarrow K^+K^-$) and NA50 
($\phi\rightarrow \mu^+\mu^-$) 
collaborations \cite{sikler99,willis99,soff99plb}.  
Contrary,  
\parbox{7.4cm}{
\psfig{figure=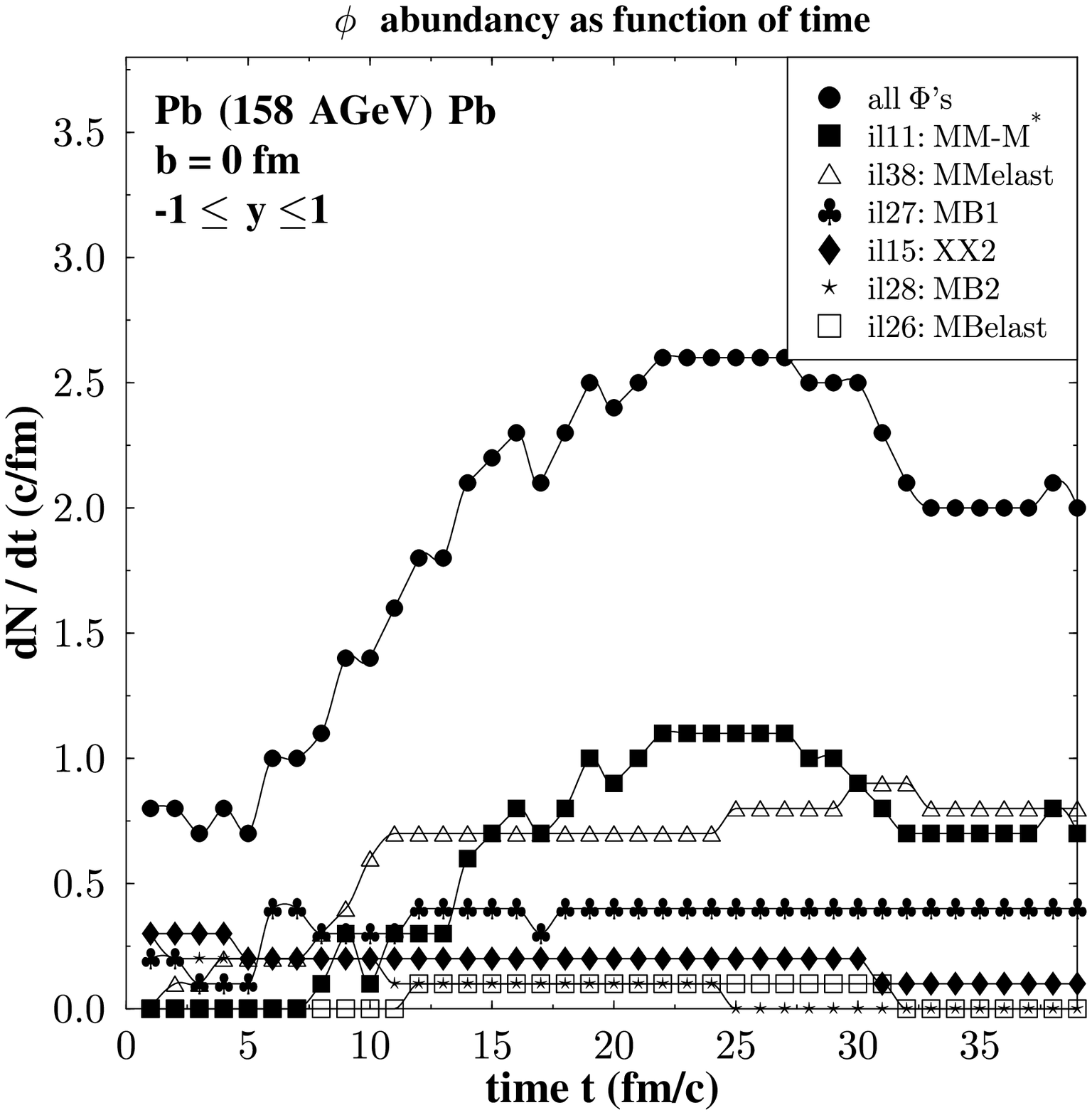,width=7.6cm}
}
\parbox{5.2cm}{{\small Figure 7: Number of $\phi$ mesons at midrapidity 
as function of time in central Pb($158\,A$GeV)Pb collisions, 
calculated with the UrQMD model (circles). The other lines 
add up to the circles and show the last processes of the 
$\phi$'s in each time step. Meson-meson collisions producing 
a $\phi$ meson (full squares), string excitation and decay producing 
$\phi$'s (club,diamond,star) and elastic collisions of $\phi$'s 
are depicted. The string excitations and elastic collisions 
are subdivided into meson-baryon collisions (MB) and meson-meson (MM) 
collisions leading to the excitation of one (MB1) or two (MB2) strings. 
(XX2) represents the string excitation by baryon-baryon (mainly first 
collisions) or 
meson-meson collisions.\\
\mbox{ }}
}
the production of $\phi$'s 
is also strongly driven by late kaon kaon scattering as 
shown in Fig.7.
In the first $5\,$fm/c $\phi$ production via string decays dominates 
while in the later stages ($10\,$fm/c$<t<20\,$fm/c) 
the resonant production via $K\overline{K}\rightarrow \phi$ 
becomes important. Approximately $40\%$ of the $\phi$'s scatter 
elastically. The strongly time-dependent $\phi$ abundancy (due 
to late resonant meson meson collisions) 
makes it questionable whether the inclusion of 
$\phi$'s in a thermodynamical 
analysis \cite{yen99,hamieh00,rafelski99} is reasonable. 

\vspace{-.5cm}
\section{Summary}
\vspace{-.2cm}
\begin{itemize}
\item
relative enhancement of strange particle production in 
central Pb+Pb collisions with respect to peripheral Pb+Pb or p+Pb collisions\\
$\longrightarrow$
enhancement grows with strangeness content of probe\\
$\longrightarrow$
UrQMD predicts enhancement due to rescattering but not 
sufficient to explain data\\
$\longrightarrow$
reduction of quark masses or enhanced string tension 
reproduces observed high yields\\ 
$\longrightarrow$
strong rapidity dependence of enhancement factors
\item
unstable (strange) particle yields ($\Lambda^*(1520),\, \phi$) are 
distorted due to 
rescattering of decay products\\
$\longrightarrow$
approximately $25\%$ of $\phi$'s and $50\%$ of $\Lambda^*$'s 
have rescattered decay products
\item
late $\phi$ production by resonant $K\overline{K}$ scattering
\end{itemize}

\vspace{-.6cm}
\ack
\vspace{-.3cm}
This work is supported by DFG, BMBF and the Alexander von Humboldt-Stiftung.  

\end{document}